# Best Practices for Managing Data Annotation Projects

Tina Tseng, Legal Analyst, Bloomberg Law

Amanda Stent, NLP Architect, Office of the CTO

Domenic Maida, Chief Data Officer, Global Data

September 2020

Version 1.0

**Bloomberg**

# Table of Contents

**XI.   After Annotation is Complete**

# I. Introduction

## Overview of an Annotation Project

*Annotation* is the labeling of data by human effort. Annotation is critical to modern machine learning, and Bloomberg has developed years of experience of annotation at scale. In the Fall of 2019, Bloomberg's Office of the CTO and its Global Data department jointly held a workshop with more than 30 experienced annotation project managers, with the goal of documenting best practices for managing annotation projects.

Before the workshop, we surveyed participants to collect project management themes they considered important for creating and managing data annotation projects. We clustered these themes into categories – one category for each section of this report. The categories correspond to common milestones of a data annotation project, as illustrated in Figure 1.

During the workshop sessions, participants discussed best practices for each category. After each workshop session, we collated the best practices identified in the session notes and constructed a survey for participants to rate the importance of each best practice to come out of that session for that category. The top-rated best practices are discussed in detail in the corresponding sections of this guide.

After the workshop, volunteer participants worked with the co-leads to author this report. We thank all of the participants for their valuable contributions. This report captures a wealth of wisdom for applied annotation projects.

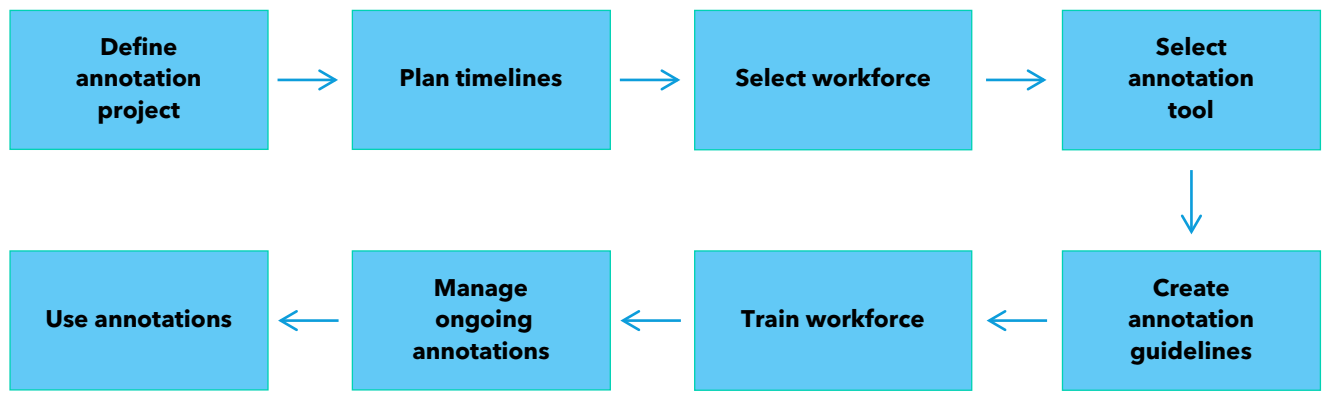

*Figure 1: Stages of a Data Annotation Project*

## II. Starting a new project

### Summary

- **Identify key stakeholders** at the start of any new annotation project.

- **Establish high-end end goals,** with input from all key stakeholders. Determining and communicating these goals will help those involved in the project understand the project's purpose and ensure that everyone is working towards achieving the same results.

- **Select methods for regular and clear communication** among the stakeholders through all stages of the project. This will maintain focus on the project's end goals. This is important whether the project involves a single team or a larger cross-team/cross-functional group.

### Identify key stakeholders

At the beginning of an annotation project, key stakeholders should be identified. Typically, there are three:

- **Project manager** determines the practical application for the project and understands its potential business impact; prioritizes features based on user needs; may have subject matter expertise.

- **Annotation project manager** is primarily responsible for the day-to-day oversight of the annotation project; selects/manages the workforce; ensures data quality and consistency; may have subject matter expertise.

- **Engineering manager** is primarily responsible for implementation of the solution for which the annotations are needed; provides input as to technical strategy and feasibility.

Collectively, these stakeholders (1) establish the project's high-level end goals, (2) plan logistics, and (3) set timelines. All stakeholders are responsible for ensuring that the organization has the right to annotate and redistribute the data, and that privacy and security requirements for the data and its annotations are considered.

Sometimes a project has only two of these stakeholders. However, any project without all three stakeholders is at risk and should be considered carefully, as all stakeholders provide a valuable perspective.

### Establish high-level end goals

With respect to high-level end goals, stakeholders should discuss what the ultimate objective is, including:

- The desired functionality, intended use cases and target users.

- The significance of the end goal, including why any existing solutions are inadequate and what the expected business impact will be.

- How success will be measured, including the milestones that need to be met and metrics achieved.

The "real-world" context for the project should drive the project's overall strategy with respect to prioritization and implementation.

#### Desired Functionality, Use Cases and Target Users

Getting into the mindset of the target user who will benefit from the project will help all stakeholders focus on a practical end goal and facilitate a strategic approach to the data annotation process. For this reason, the product manager should give all stakeholders a product overview and explain the desired functionality from the target user's perspective. Stakeholders should discuss what the target user's needs are and how the project will address those needs. Background on the industry, user personas, and workflow should be provided. It may also be useful for the stakeholders to create wireframes and talk to the target users directly to better understand their expectations and desires.

**Significance of the End Goal**

In addition to establishing the desired functionality, stakeholders should discuss any existing alternatives and the expected business impact of the project. Discussing current solutions to the problem being addressed will help the stakeholders understand what can be done better, how those improvements can be achieved, and the value of those improvements. In addition, discussing the potential business impact of the end goal will give the stakeholders an understanding of the overall importance of the project and how that can affect the available resources and ultimate timeline for the project.

**Measuring Success**

Stakeholders should agree on how to determine whether the project has been successful. For example, they should agree on what the project's milestones are, such as requirements for a minimum viable product. Stakeholders should also discuss any applicable metrics, like precision and recall, that will be used for performance evaluation. To be meaningful, metrics should always be aligned with user expectations. Making milestones and metrics requirements explicit at the outset of the project will help manage stakeholders' expectations and inform the design of the annotation project.

## Select Methods for Regular and Clear Communication

The key stakeholders in an annotation project should agree on methods for regular and clear communication.

- Methods that work well include Scrum, Kanban or the Dynamic Systems Development Method.

- At a minimum, the stakeholders should establish a shared project roadmap, backlog of issues, and regular meetings for the annotation project.

The annotation project manager should have the authority to manage the project roadmap and backlog, as well as to determine meeting agendas because he or she will best understand any issues that arise.

# III.  Defining the Annotation Project

## Summary

Defining the parameters of the annotation project should be a collaborative effort among stakeholders to ensure that the project's design ultimately leads to the desired high-level end goals:

- **Identify the data to be annotated, and determine the types of annotations needed.**

- **Decide how best to collect annotations**, which may entail subject matter experts, partial automation, and/or microtasking.

- **Consider the budget, available resources, and timeline.**

- **Consider how the annotations will be used**, including whether the annotations will be used for automation (e.g., to train a machine learning model) or passed on directly to the target users.

## Identify Data and Determine the Types of Annotations Needed

The stakeholders should identify the characteristics of the data to be annotated, how the data will be sampled, and the likely annotations that need to be captured.

Exploring the data is an important initial step of the project definition process because understanding the data's features, limitations, patterns and edge cases will allow those involved in the project to make informed decisions about the types of annotations required.

Once the stakeholders understand the data, they should decide how to sample for annotation and whether any pre-processing of the data is needed before annotations are applied.

They should then make an initial version of the label set for the annotations.

## Decide How Best to Collect Annotations

Once the stakeholders have explored the data set and determined the types of annotations required, they should consider how best to collect annotations.

- Some types of annotation require subject matter experts who can look at all of the data (or large subsets of the data) at a single time.

- Other types of annotation can be achieved via partial automation through rules, accompanied by some human labeling ("weak supervision").

- When a project calls for a large number of annotations involving different types of labels, stakeholders should consider whether executing a series of distinct microtasks would be a viable approach. A "microtask" is the smallest unit of work that requires human intelligence to complete.

Careful attention should be paid at this stage; if a project can be broken down so that it is amenable to microtasking, this can reduce the time and expense of annotation, as well as lead to higher quality and potentially reusable data. Microtasking also allows the development team to more easily identify pain points and iterate.

For annotation projects that are complex and require subject matter expertise, one or more of the following may be necessary:

- Break down the project into subprojects that are simpler and/or do not require subject matter expertise. Many complex projects can be broken down into simpler microprojects, at least some of which can be assigned to annotators without specialized expertise.

- Provide background information and training on the data set/subject matter as part of annotator training, and evaluate annotators' learning. This may add to the timeline for the project, but may have only a minimal impact on the budget.

- Select an expert workforce. Finding a workforce with subject matter expertise may take more time and be much more expensive than using a workforce without specialized expertise.

## Consider Budget, Resource Availability, and Timeline

Budget and resource availability will necessarily affect the way the project is defined because it may restrict the type of workforce that can be used for the project. Depending on the nature and complexity of the data set, the type of workforce may affect the number and types of annotations that can be produced. Similarly, timeline considerations may affect the way that the project is defined because a limited time frame may also limit the number and types of annotations.

## Consider How the Annotations Will Be Used

The nature of the data set and the end goal of the project should inform how the annotations will ultimately be leveraged. Sometimes, the annotations will be passed directly on to end users; in this case, stakeholders should carefully consider whether automation might help. Stakeholders should consider a machine learning model when the goal of the project is to make future predictions on new data. Stakeholders should consider a rules-based model when the goal of the project is to capture human-identifiable patterns in the data.

# IV. Managing Timelines

## Summary

It is important for all stakeholders to be involved in constructing timelines for the project. Timelines communicate expectations, constraints and dependencies, which can have a large impact on budget and project success.

- Stakeholders should **collaborate on project timelines** so that expectations can be managed and perspectives balanced.

- Stakeholders should **ensure timelines are clear and detailed**, including enough information about projects and their dependencies as well as key milestones.

- When establishing timelines, it is important to **include time for creating guidelines and training the workforce**.

## Collaborate on Project Timelines

Project timelines should take into account the project's priority and high-level end goals, as well as logistical and resource constraints. All stakeholders should collaborate on an initial project timeline with brief written descriptions of the project's major milestones. When reviewing project timelines, each stakeholder should provide input based on their unique perspectives:

- The annotation project manager should ensure that the timelines are realistic based on their knowledge of the data set, the project's complexity, and the type of workforce that is available to do the annotations. Any issues or uncertainties related to the data and the annotation process should be communicated to all stakeholders and documented as risks, where applicable.

- The engineering manager should ensure that the timelines are aligned with the development requirements for the product solution. If there are any changes in expectations as the project progresses, such as increased annotation volume requirements, these should be communicated to the other stakeholders.

- The product manager should ensure that the timelines take into account any applicable product requirements, user expectations, and deadlines. Since the product team is often the least involved with the logistics of the annotation project, it is especially important to ensure that they have a strong understanding of the project's complexity in order to manage their expectations.

Although the annotation project manager is often tasked with implementing and maintaining the timeline for the annotation process, all stakeholders should be aware of how the annotation project's complexity may affect the project's timeline. To manage expectations and foster strong communication throughout the lifecycle of the project, any timeline changes due to unforeseen complications or limitations should be communicated to all stakeholders as soon as possible.

If the team has adopted an Agile process as recommended in **Section II**, timelines can be discussed during regular team meetings, and realized on the project's backlog. Written communications about the timelines serve as an important record of stakeholders' constraints and expectations.

## Ensure Timelines are Clear and Detailed

After the initial project meetings, stakeholders should create a more detailed set of timelines. These timelines should include milestones for:

- data preparation and annotation;

- solution implementation; and

- product development.

Dependencies between milestones (e.g., that a certain amount of annotated data must be available before solution implementation starts) should be clearly identified. Each milestone should have a description and examples. In addition, each milestone should be timeboxed (e.g., in terms of sprints). The timebox for a milestone may be impacted by the complexity of the project and amount of data to be annotated. For example, if a large number of annotations is needed, then multiple sprints will be necessary. The same holds true for the evaluation of a machine learning model against a test set: the larger the test set, the more work will be needed. The annotation project and engineering managers should be transparent with the product manager about timeline requirements, based on previous projects or thoughtful estimates. New annotation project managers should consult with experienced ones about the content of their timelines.

After the timelines are drafted, it is important to ensure that the layout of the timelines document itself enables easy digestion and comprehension by the stakeholders. If there are many timelines, it may make sense to separate the milestones across different Program Increments to ensure proper time is planned (e.g., one or two milestones per sprint).

## Include Time for Creating Guidelines and Training the Workforce

It is important to include time for drafting and norming annotation guidelines, and for selecting and training the workforce. Poor annotation guidelines, or a poorly prepared workforce, may result in wasted effort and an increase in back-and-forth with the annotators that will add more time and avoidable uncertainty to the annotation process. If the annotation project is complex or requires a large number of annotations, this should be communicated to all stakeholders in order to manage timeline expectations properly, as more extensive guidelines and training may be needed.

**Section V** describes the different types of workforces and the impact that the workforce selection can have on timelines and budget. For example, the Crowd is often ideal for annotation projects with shorter timelines and public data. Other factors, such as project complexity and volume, should also be considered.

# V. Selecting the Workforce Type

## Summary

Selecting the workforce used to complete the annotation project requires a good understanding of the data's privacy and security requirements, the project's complexity, and budgetary constraints. The annotation project manager should always have input on the choice of workforce; this decision should never be made solely by the product or engineering managers.

- It is important to **know the workforce options**, as each workforce type has benefits and limitations.

- It is also important to **understand the constraints on workforce selection**. Data privacy restrictions, project complexity, annotation volume, annotation complexity, required subject matter expertise, availability to supervise and provide quality assurance, as well as the project's lead time, are all factors that should be considered when selecting a workforce.

## Know the Workforce Options

To ensure that the most appropriate workforce is selected for an annotation project, stakeholders should explore different workforce options to weigh their benefits and limitations. There are primarily three types of workforces:

### The Crowd

The Crowd is the most cost-effective option and requires the least amount of lead time. They are great at performing simple projects that require minimal skill and training. As such, microtasking is often ideal for ensuring quality of crowdsourced annotation. The training that the Crowd receives is communicated through written instructions provided on the crowdsourcing platform on which the project is hosted, and there is limited ability to supervise and communicate with individual annotators. Depending on the crowdsourcing platform that is used, gold tasks, double-blind validation, and annotator random sampling may be employed to perform quality assurance on the Crowd's work.

### Vendors

Vendors are relatively low-cost workers that can handle basic and potentially skilled work. Subject to confidentiality/non-disclosure agreements, vendors can access an organization's internal tools and certain non-public data. In addition, vendors can sometimes support more complex annotation projects because their work can be closely monitored. For these reasons, vendors are often chosen when the data must be kept private, the annotation task is more complex, or subject matter expertise is required. Hiring and onboarding vendors requires more lead time than for the Crowd. In addition, depending on the location of the vendors, their work may be done outside of stakeholders' normal business hours, which may present communication and supervision challenges.

When engaging vendors for an annotation project, consideration should be given to the pricing structure of the engagement, as financial incentives may affect annotator behavior. Transactional pricing may encourage faster annotation throughput, while flat-rate pricing with quality metrics may encourage higher quality annotations.

### Full-Time Employees

Full-time employees are best for annotation projects that are complex, private, require subject matter expertise, or where no dedicated workforce budget for the project exists. They are the only option when dealing with data that contains personally identifiable information or other information restricted by law. Full-time employees still require training and supervision, with the project managed by an annotation lead.

## Understand the Constraints on Workforce Selection

### Data Privacy Restrictions

The stakeholders should carefully consider the type of data being annotated. Intellectual property, data privacy, and security concerns will impact the workforce type and annotation platforms that can be leveraged for an annotation project. Always err on the side of caution and reach out for official guidance (e.g., from an organization's Internal Review Board or Legal/Compliance department/s) if there is any uncertainty.

### Project Complexity

Project complexity should be considered when selecting a workforce. One should understand the volume, complexity, and subject matter expertise required for the annotation, as well as the amount of supervision, quality assurance, and lead time associated with managing the project itself. These factors will impact cost and, in turn, should be weighed against budgetary considerations. Complex projects can often benefit from a pilot project that can efficiently determine the efficacy of the project at low cost. New annotation project managers should consult with experienced ones about how their project's complexity may impact their workforce selection.

### Annotation Volume

Annotation volume is the number of annotations required to sustain the project. Projects that require a large volume of annotations may need a scalable workforce such as the Crowd or vendors. If the project is unsuitable for these workforce types due to privacy restrictions, adjusting the timeline or determining how to build a minimum viable product with a smaller number of annotations may be required.

### Annotation Complexity

Annotation complexity refers to the level of precision and number/type of considerations that the workforce needs to take into account to generate accurate annotations. An example of a consideration that could impact annotation complexity is a determination of which textual sub-components should be annotated to indicate the salience of a data point. Complex annotations are best suited for vendors or full-time employees because they can be provided with extensive training, supervision and feedback.

### Required Subject Matter Expertise

Subject matter expertise refers to knowledge that is more comprehensive than documentation can provide. It is sometimes necessary for the workforce to have certain subject matter expertise already when executing an annotation project, because the level of training required to transfer this knowledge can be too time-consuming and costly. Although limiting the potential pool of crowdworkers to certain demographics may help annotation project managers recruit contributors who have the required subject matter expertise, vendors and full-time employees are likely better suited than the Crowd for these projects because it may be difficult to recruit a sufficient number of qualified crowdworkers in the desired timeframe.

**Ability to Supervise and Provide Quality Assurance**

No matter the volume, annotation complexity, and subject matter expertise required for an annotation project, all projects require some supervision and quality assurance. When selecting a workforce, the annotation project manager's ability to supervise the annotation process on a consistent and potentially real-time basis should be a consideration, as vendors and full-time employees typically require more hands-on supervision than the Crowd.

**Project Lead Time**

Project lead time should be considered when selecting a workforce because limited lead time may impact the feasibility of engaging a vendor. As discussed above, lead time is often required to establish the vendor relationship: selecting which vendor to use, establishing a contract, hiring the annotators for the project, and ensuring that the vendor has access to the necessary tools to do the annotations can take a significant amount of time. On the other hand, utilizing the Crowd requires relatively little lead time because a Crowd job can be deployed as quickly as the annotation project is designed and data is loaded onto the crowdsourcing platform. Full-time employees may also be a viable option if lead time is low, assuming their priorities are aligned with the project.

# VI. Selecting an Annotation Tool

## Summary

When selecting an annotation tool, there are several factors that stakeholders should consider and balance.

- Engineering and annotation project managers should endeavor to **use standard infrastructure**. Selection of an existing and supported annotation platform or tool should be preferred over implementation of a new tool or ad hoc workflow.

- The primary drivers for tool selection should be to **ensure data privacy and access by the workforce**.

- The stakeholders should also **evaluate technical back-end requirements**.

- It is also important to consider how the tool will **facilitate project management needs**.

- Stakeholders should also **establish desired UI/UX features** for the annotation tool and project interface.

## Use Standard Infrastructure

Implementation and maintenance of new annotation tools leads to technical debt over time and potential governance issues. Whenever possible, choose an existing annotation platform or tool that is already supported by the organization.

## Ensure Data Privacy and Access by the Workforce

As outlined in the previous section, intellectual property, data privacy, and security considerations impact the workforce type and which annotation tools can be used for an annotation project. Stakeholders should agree on a tool that can be used in the overall project workflow and that the chosen workforce can access.

## Evaluate Technical Back-End Requirements

At a minimum, the selected tool should enable efficient upload, download, storage, and processing of data. The tool should also support standardized and easy-to-use data formats like JSON and XML.

## Facilitate Project Management Needs

Project management functionalities should be built into the tool to help annotation project managers monitor the status of the project, communicate with the workforce, identify issues with the annotation process, and control for quality. Specifically, the selected tool should accommodate annotation project managers' need to:

- convey written guidelines and updates to the workforce;

- control work queues, including priority/order;

- access audit logs for work items;

- perform quality assurance, including leveraging annotation redundancy and "gold" tasks;

- provide feedback to the workforce; and

- track quality, productivity and timeliness metrics.

## Establish Desired UI/UX Features

The selected tool should have an interface that can be customized to the requirements of the annotation project in order to give the workforce the ability to execute the annotations in an efficient and accurate manner. Stakeholders should ensure that the UI/UX features of the tool do not bias the workforce, thereby impacting the validity or reliability of the annotations. The workforce should have the ability to easily:

- access the written guidelines and updates for the project;

- read/view the relevant data in the work item;

- add/edit/delete annotations;

- navigate through each work item/work queue, including revisiting previously-submitted work items, if that is desirable; and

- flag work items with potentially corrupt or unsuitable data, if that is desirable

# VII. What to Include in Annotation Guidelines

## Summary

To ensure clear and consistent communication with the workforce, annotation guidelines should serve as the main, if not sole, reference point for the annotation project. As such, they should be easily accessed by the workforce on the annotation tool during the annotation process and contain all of the information needed to complete the project. When creating a new project, it may be helpful to review previous projects as baselines or sources of inspiration.

- When writing annotation guidelines, it is important to **keep complexity and length in mind**. The complexity of the guidelines will necessarily be influenced by the complexity of the annotation project and type of workforce performing the annotations.

- Annotation guidelines should **include both tool and annotation instructions**.

- Annotation guidelines should also **illustrate each label with examples**. Examples will help the workforce understand the data scenarios more easily than lengthy explanations.

- Annotation project managers should **consider including the end goal** or downstream objective in the annotation guidelines to provide context and motivation to the workforce.

- Annotation guidelines should always **be consistent with other written documentation** about the project to ensure the workforce does not have conflicting or confusing guidance.

## Keep Complexity and Length in Mind

The type of workforce selected for the project will impact the complexity and length of the annotation guidelines. Guidelines for Crowd projects should be succinct and straightforward. Complicated reference material may result in difficulties recruiting contributors, given that crowdworkers are paid per work item.

Guidelines for vendors and full-time employees can be more extensive if the nature of the project requires it. For example, if domain knowledge is needed to execute the annotations accurately, these guidelines should provide the necessary background information on the data set/subject matter to help the workforce understand the project. This background information should be provided as part of the written guidelines rather than as a separate oral presentation. This ensures that communication of the information is clear and consistent.

When the guidelines are relatively complicated or lengthy, the information within the document should be searchable and discoverable to help the workforce quickly find the relevant guidance during the annotation process. For example, within the guidelines document itself, page numbers and a table of contents with hyperlinks can be useful for navigation. In addition, depending on the annotation tool, a summary of the guidelines with representative examples can be embedded in each work item so that guidance can be viewed simultaneously alongside the data to be annotated.

## Include Both Tool and Annotation Instructions

The workforce should be provided with functional instructions, explaining how to use the interface of the annotation tool, as well as substantive instructions describing how to execute the annotations based on the labels properly. Both types of instructions could be provided in a single document or in separate ones, depending on complexity and length.

Annotation interface instructions should detail how to operate and troubleshoot the annotation tool. Some fundamental concepts to cover are: accessing the annotation guidelines and work queues; interacting with work items (saving, submitting, and skipping/revisiting, if applicable); and interacting with annotations (adding, modifying, and deleting). These concepts should be conveyed in a step-by-step format so that the chronology of the workflow within the tool can be understood and followed.

Substantive annotation project instructions should include descriptions of each label and explain how they should be applied to the data set. The paramount concern when drafting these instructions is whether the labels are being adequately defined and explained so that the annotation of the data set will be as accurate and consistent as possible. To that end, the guidelines should state if more than one label and/or differing annotations could be valid. Additionally, the guidelines should state whether annotations can be discussed among annotators; if annotation redundancy and consensus are being leveraged for quality assurance, it is necessary to state in the guidelines that annotators cannot discuss their work items with one another and that they should not ask for guidance from annotation project managers beyond the written guidelines.

## Illustrate Labels with Examples

Examples are crucial for illustrating labels so that the workforce is aware of common and representative data scenarios, as well as outliers. Often, examples are more effective at communicating concepts than lengthy explanations. Where possible, examples should consist of up-to-date screenshots of an annotated work item with descriptions that highlight features to which the workforce should pay particular attention. In addition to common and edge cases, examples of counter cases may be useful when there is ambiguity or a perceived overlap between labels. Hyperlinks within the annotation guidelines can be used to compare and contrast different examples.

The number of examples included in the guidelines will depend on the annotation complexity. Unintuitive and ambiguous labels will require more examples than relatively straightforward labels.

## Consider Including the End Goal

For some annotation projects, it may be useful to explain the project's end goal in the guidelines document so that the workforce has context for why the annotations are being collected. Understanding the project's purpose may help the workforce better understand the requirements of the project, as well as provide motivation for executing what may otherwise be perceived as a tedious undertaking.

## Maintain Consistency with Other Written Documentation

When vendors are engaged as the workforce, they may create and maintain detailed notes in addition to the provided guidelines in order to help them on-board and train new annotators. Although it is preferable for the provided guidelines to serve as the sole written reference point for an annotation project's workforce, this may be a viable approach for vendors who have been working on a project for an extended period of time with a proven track record of high-quality annotations. Any written documentation that is being used in addition to the guidelines must be reviewed by annotation project managers to confirm their accuracy and consistency.

# VIII. Creating, Testing and Maintaining Annotation Guidelines

## Summary

- It is best to **designate a single person to "own" the guidelines** so that the document maintains internal consistency.

- Annotation project managers should **employ a limited pilot period to test the guidelines** to confirm that they produce the desired output. Testers can annotate a small number of items and then collectively review them to assess whether changes to the guidelines should be considered.

- When possible, it is best to **display the guidelines in the annotation tool** so that the workforce can easily refer to them as part of their workflow.

- Annotation project managers should **communicate changes to the guidelines to the workforce in writing** so that all annotators receive the same information simultaneously and continue to generate consistent annotations. All stakeholders should carefully consider and explicitly approve any modifications to the guidelines that affect the validity of previously-annotated data, as these modifications will likely impact the project's cost and timeline.

## Designate a Single Person to "Own" the Guidelines

One team (ideally, one individual) should own the drafting and updating of the guidelines. This helps ensure internal consistency (i.e., no contradictory content) and continued readability. The annotation project manager is typically best-suited to author the annotation guidelines, given his or her understanding of the data set, end goal, and technical requirements. Other factors to consider when selecting an author include whether the individual has an understanding of the data set's broader applications outside of the annotation project, and whether the individual has capacity to provide continued support of the guidelines after the annotation project begins and before the project matures.

Before drafting the guidelines, it may be helpful for stakeholders to "reverse-shadow" the author while he or she annotates a few work items to ensure that the author understands the annotation process and can competently draft the guidelines.

While one person should be responsible for drafting and updating the guidelines, the final guidelines should always be a collaboration among all stakeholders. The annotation project manager should ensure that the guidelines are clear and consistent based on his or her knowledge of the data set. Engineering managers should ensure that the guidelines are aligned with downstream needs and metrics. Product managers should ensure that the guidelines take into account any applicable product requirements or user expectations.

## Employ a Limited Pilot Period to Test the Guidelines

Before annotation collection officially begins, the guidelines should be "tested" to ensure they are easily understood, produce consistent annotations, and result in a usable output. Tester feedback, as well as testing output, can provide insight as to whether or not the guidelines need to be changed. However, it is critical to restrict the testing period to a limited duration to prevent project delays.

Testers may either be the project stakeholders or part of the workforce that will be doing the annotations. If additional training materials outside of the guidelines (such as a training session) will be used for the actual workforce, the testers should be provided with the same.

There are two ways to gain insights from the testing period:

**Annotation redundancy.** If the annotation tool supports annotation redundancy, testers should annotate the same work items and should not collaborate. For annotations with consensus, check that the desired output has been achieved. For conflicting annotations, speak with each annotator to discover how the guidelines were interpreted and applied. This can help the stakeholders identify where the guidelines could be clarified for greater annotation consistency.

**Individual annotation review.** While this approach is less efficient, individual annotation review allows testing in cases where there is only one available annotator or the annotation tool does not support annotation redundancy. As above, speak directly with the annotator to identify how the guidelines were interpreted and applied, especially for incorrect annotations or inconsistent behavior.

## Display the Guidelines in the Annotation Tool

There are several factors to consider when choosing a platform for annotation guideline management:

- documentation support in the annotation tool;

- privacy concerns; and

- version control.

Annotation guidelines for the Crowd should always be displayed in the annotation tool because crowdworkers are unlikely to have access to other platforms or repositories. Guidelines for vendors and full-time employees do not necessarily need to be displayed in the annotation tool, as long as the selected platform makes the documentation consistently available and easily retrieved. Nevertheless, it is ideal for the guidelines to be displayed in the annotation tool for all workforce types so that annotators are encouraged to regularly refer to the document during the annotation process and verify that their annotations conform to the rules and examples contained therein.

When determining where to *store* the guidelines, it is imperative to consider the privacy of the underlying data and annotations. Guidelines that relate to private data may be restricted to specific internal repositories. On the other hand, guidelines relating to public data can likely be hosted on a larger variety of platforms.

Another important consideration when storing guidelines is whether the hosting platform supports document versioning so that changes to the guidelines can be tracked over time. Annotators should never store local copies of the guidelines outside of the chosen document management system, as this can result in outdated guidelines being used and inconsistency among annotators. Hosting the guidelines on the annotation tool or in a shared repository ensures that the workforce is referring to the same version of the document at the same point in time. If any changes or updates are made to the guidelines during the annotation process, the workforce, as well as all of the stakeholders, should be able to track how the guidelines evolved and when any changes or updates went into effect.

If the guidelines can only be shared with the workforce via emails and attachments, care should be taken to reduce potential issues by:

- sending the attachment in an uneditable format, such as PDF;

- clearly indicating the version number and effective date at the top of the document and in the file name;

- regularly verifying with the workforce that the current effective guidelines are being used; and

- only using this approach if the annotators have previously demonstrated the ability to self-manage their documentation practices.

## Communicate Changes to the Guidelines

As the annotation project proceeds, edge cases and other unexpected data scenarios will likely be encountered. As these cases may impact annotation consistency, it is critical to address them quickly, in a controlled manner. An update or modification to the annotation guidelines may be warranted when there appears to be widespread confusion regarding label application or how specific annotation scenarios should be handled. To ensure internal consistency, the same individual who authored the initial guidelines should incorporate any updates/modifications to the document.

**Updates** are clarifications to the guidelines that do not change or contradict the existing guidelines. Because updates simply provide further explanation or additional examples consistent with the existing guidance, they should not affect the validity of any previously-annotated data. Updates should be circulated to all stakeholders prior to their communication to the workforce so that any objections or unforeseen downstream impacts can be raised.

**Modifications** are changes to the guidelines that impact the validity of previously-annotated data because they alter or even contradict existing guidance. These changes should only be considered if the current guidelines are not producing the desired output (e.g., in terms of annotation structure for a machine learning model, or specificity of the attributes for extraction scope). *Modifications that change or contradict any existing labels or rules should be explicitly discussed and approved by all stakeholders because reviewing and remediating or discarding previous annotations will increase the duration, effort and cost of the project.*

Once any updates or modifications to the guidelines have been incorporated, the changes should be communicated to the workforce in writing, along with the effective date. This ensures that all annotators receive the changes simultaneously, which promotes annotation consistency. The written communication should include an overview of what the changes are and where the changes can be found in the body of the guidelines document so that the workforce immediately knows which sections to review more carefully.

It is important to track the effective date of updates and modifications, as stakeholders should identify which annotations may need to be remediated in order to conform to the final guidelines. The guidelines document should clearly indicate the effective date in both the file name, as well as the document title.

# IX.  Staffing and Training the Workforce

## Summary

How the annotators are trained for an annotation project depends on the nature of the project, the project's time frame, preferences, and the resources of the individual or group managing the workforce. However, training is an essential part of the annotation process to ensure that the annotators adequately understand the project and produce annotations that are both valid (accurate) and reliable (consistent). As mentioned in **Section IV**, the timeline for the project should take into account a training period, the length of which will depend on the workforce, complexity of the project, and quality assurance method employed.

- It is important to **use the guidelines for training** the workforce so that the guidance remains consistent, no matter when the annotators are onboarded. If the guidelines need to be supplemented, the additional training should be provided in writing or recorded video.

- **Real-time questions from the workforce should be encouraged** during the training period so that any confusion or misunderstanding of the guidelines can be mitigated as soon as possible. If real-time questions cannot be accommodated, annotation project managers should have debriefs with the annotators to discuss any issues.

- It is essential to **assess quality during the training period**. Quality assurance methods can be used to determine feedback for the workforce and evaluate annotator performance.

- Annotation project managers should **provide written feedback during the training period** so that the workforce can refer to it. If necessary, annotation project managers should also update the project's guidelines based on tester feedback and/or testing output.

## Use the Guidelines for Training

Annotation project managers should use the project's written guidelines as their main tool to train the workforce. As described above, the guidelines should clearly detail the parameters of the project, including a description of the annotation platform's functionality, chronological annotation instructions, and explanations of examples representative of the data set and labels. Making the guidelines the workforce's principal source of guidance will ensure that all annotators receive the same training no matter when they are onboarded to the project. In addition, consolidating all of the guidance into a single document will make it administratively easier for the annotation project manager to communicate any updates or revisions to the initial training as the project progresses.

If it is impractical to solely rely on the guidelines to train the workforce, annotation project managers should endeavor to provide any additional training in writing or recorded video. This way, there is little risk of inconsistency in training for annotators who were onboarded at different times, and there is a dated record of all of the guidance given to the workforce for the project.

## Encourage Real-Time Questions

During the training period, annotators should be encouraged to ask questions if they are uncertain about the annotation tool's functionality or application of the project's guidelines so that any confusion or misunderstanding can be addressed quickly.

One option is to have a virtual chat room to facilitate real-time questions. If real-time questions are allowed, it is important to support collective communication with the workforce and to keep a written record of all communications. This will help maintain consistency in responses and allow annotation project managers to track any changes to the workforce's annotations that are due to changes in guidance.

Real-time questions may need to be limited in certain circumstances, such as:

- when the project requires subjective judgment calls (and additional guidance may taint the results);

- when "gold" tasks or annotation redundancy is being leveraged to assess the quality of the annotations produced (see below); and

- when the annotators work outside of the annotation project managers' normal business hours.

In those cases, debriefs with the annotators after they have completed a certain number of work items may be used. Debriefs are scheduled discussions with either individual annotators or the workforce as a group where questions about the guidelines and specific examples may be raised. Debriefs can be conducted in-person, over video conference, or via email. However, if an in-person or video conference is used for debriefs, it is best for annotation project managers to memorialize what was discussed in writing so there is a dated record of any guidance provided.

## Assess Quality During the Training Period

The quality of the annotations that the workforce produces must be analyzed as part of the training process so that feedback to the workforce can be provided and the performance of individual annotators, as well as the workforce as a whole, can be evaluated. Methods of assessing quality include:

- "gold" tasks;

- annotation redundancy with targeted quality assurance (QA) or debrief; and

- random QA.

### "Gold" tasks

"Gold" tasks are work items that can be compared to "answer keys" which have been previously prepared by annotation project managers. Using "gold" tasks as a way to train the workforce can be ideal for objective annotation projects where there are strictly defined ways to annotate. "Gold" tasks shorten the feedback loop between annotation project managers and the workforce because the "correct" annotations for those work items have been determined in advance and can be quickly communicated to the annotators. However, creating "gold" tasks for training requires a significant amount of lead time for data exploration so that annotation project managers can ensure that there are a sufficient number and variety of "gold" tasks to be representative of the actual data set.

### Annotation Redundancy with Targeted QA

Annotation redundancy requires more than one annotator to annotate the same work items independently so that their annotations can be compared. Annotation project managers can then perform targeted QA by prioritizing work items with differing annotations (i.e., "disagreement" or "lack of consensus") for review because those work items are most likely to have accuracy or consistency issues related to annotator confusion or ambiguity in the guidelines. Although those work items should be prioritized, a percentage of the work items with identical annotations (i.e., "agreement" or "consensus") should also be reviewed by annotation project managers to ensure that those annotations are valid and comply with the project's guidelines. Although annotation redundancy with targeted QA does not require advance preparation by annotation project managers or delay the commencement of the actual project, it does involve a longer feedback loop because feedback to the workforce is delayed while the annotators complete the work items and those work items are reviewed. In addition, the annotation tool must have the ability to accommodate annotation redundancy.

### Annotation Redundancy with Debrief

Annotation redundancy can be followed by debriefs with the annotators in lieu of targeted QA. This method of training the workforce may be more desirable for subjective annotation projects where there is a larger range of acceptable annotations. During debriefs, annotation project managers should discuss with the annotators work items where the annotators agreed, as well as ones where they disagreed, to determine how the annotators applied the objective factors of the project's guidelines. Although debriefs provide annotation project managers with greater insight into the annotators' thought processes so that more effective feedback can be provided, they also take more time than simply providing the annotators with copies of erroneous work items with explanations of the correct annotations. As mentioned above, debriefs should always be followed by written summaries of any guidance that was provided during the discussions so that there is a record of all training that the workforce received.

### Random QA

Random QA entails randomly sampling work items from each annotator to scrutinize for quality. This method of training the workforce can be used when the annotation tool does not support annotation redundancy and when the annotation project calls for a large quantity of data to be reviewed within a limited period of time. Although random QA does not require lead time for advance preparation or delay the annotation of the actual data set, it also does not quickly identify likely errors or areas of confusion.

| Method | Benefits | Considerations |
|---|---|---|
| "Gold" tasks | • Prompt evaluation<br>• Short feedback loop | • Delays annotation of actual data set<br>• Requires lead time to ensure that "gold" tasks are representative of the data set |
| Annotation redundancy and targeted QA | • Prioritizes likely errors and areas of confusion<br>• Does not delay annotation of actual data set<br>• No lead time required | • Slower evaluation<br>• Longer feedback loop<br>• Reduced productivity, due to overlapping work items<br>• Annotation tool must accommodate annotators completing the same work items simultaneously |
| Annotation redundancy and debrief | • Can be effective for subjective projects<br>• Prioritizes likely errors and areas of confusion<br>• Does not delay annotation of actual data set<br>• No lead time required | • Slowest evaluation<br>• Longest feedback loop<br>• Reduced productivity, due to overlapping work items<br>• Annotation tool must accommodate annotators completing the same work items simultaneously |
| Random QA | • No impact on productivity<br>• Does not delay annotation of actual data set<br>• No lead time required<br>• Does not require annotation tool to accommodate annotators completing the same work items simultaneously | • Slower evaluation<br>• Longer feedback loop<br>• Does not prioritize likely errors and areas of confusion |

## Provide Written Feedback During the Training Period

No matter how quality is assessed during the training period, the workforce should be provided with written feedback for erroneous work items so that they can produce more accurate and consistent annotations going forward. Feedback should consist of copies of the erroneous work items, as well as clear, written explanations of the errors, with references to specific rules or examples from the guidelines and how they are applicable. Thus, feedback should always be consistent with the guidelines. If a specific data scenario calls for feedback that is not covered by or departs from the guidelines, consider whether an update or modification to the guidelines is warranted (see **Section VIII**). It is always easier and more cost-effective to modify guidelines during training than midway through a project.

# X. Managing the Annotation Process

## Summary

- Stakeholders should **review the annotation tool and workforce** to confirm the choices satisfy both policy and product requirements.

- It is important to **set reasonable quality, timeliness, and productivity targets** for the workforce based on the pilot and training periods, so that stakeholders' expectations are clearly defined and managed. These targets can be adjusted as the annotation project progresses.

- Stakeholders should agree on and **implement a quality assurance process** to ensure that the workforce produces valid and reliable annotations.

- Annotation project managers should **explicitly address annotator collaboration** with the workforce, as collaboration may be beneficial for some projects, while tainting the results of others.

- Annotation project managers should **provide the workforce with written feedback for erroneous work items** as part of the quality assurance process.

- When appropriate, annotation project managers should also **promptly respond to questions from the workforce in writing**.

- Annotation project managers should **track quality, timeliness, and productivity statistics**, which can inform the quality assurance process, as well as strategic decisions about the project and workforce.

- Annotation project managers should **ensure technical support for the annotation tool** throughout the annotation process to prevent project delays

## Review the Annotation Tool and the Workforce

Now is the time to conduct a final review of the annotation tool and workforce selections. Check that they satisfy policy constraints  and conform to product requirements (e.g., that the vendor Statement of Work (if any) matches the project plans, that the annotators have the required expertise, and that the process for communication with the annotators is clear).

## Set Realistic Quality, Timeliness and Productivity Targets

It is important to set reasonable quality, timeliness (speed to do one work item), and productivity (number of work items to be completed) targets for the annotation project based on the pilot and training periods. Using data from the pilot period when the guidelines were "tested" and the training period with the actual workforce as baselines for these targets will help establish realistic goals, because those periods most closely resemble the conditions of the actual workforce and annotation project.

Productivity and timeliness targets should take into account the complexity of the data, guidelines, and project, as well as the mechanical process of operating the annotation tool. These targets may need to be re-evaluated and adjusted periodically as the annotation project progresses, given that they should always take into account any downtime that may be attributed to technical issues or delayed responses from annotation project managers. They should also account for a learning curve, where the annotators will likely become more efficient as they become more accustomed to the tool and guidelines.

Quality targets should take into account the nature of the annotation project, as well as the end use of the data. For example, objective projects may have more specific quality targets than projects that require subjective judgment calls, because objective projects have a narrower range of acceptable annotations. In addition, the end use of the data may affect the quality target due to downstream technical requirements, such as the needs of a machine learning model.

## Implement a Quality Assurance Process

After the training period is over, stakeholders should discuss how the quality of the annotations should be assessed during the actual project and how often. Just as in the training period (see **Section IX**), the validity (accuracy) and reliability (consistency) of the annotations should be verified throughout the annotation process. Annotation quality can be assessed in a few ways:

- "gold" tasks,

- annotation redundancy with targeted QA and/or

- random QA.

During the actual annotation process, **"gold" tasks** should be interspersed among regular work items in the annotators' work queues so that the annotators are "tested" periodically. It is recommended that at least 5% of work items be "gold" tasks. This percentage should be higher if the annotators are inexperienced (to help in continuous training) or if there are many work items that will require no annotation (to check that annotators are paying attention). "Gold" tasks should be indistinguishable from regular work items so that the annotators do not know that those work items will be audited automatically for annotation quality. In addition, the inventory of "gold" tasks should be updated regularly so that the annotators continue to see new ones as they proceed through their work queues.

**Annotation redundancy with targeted QA** is another way to assess quality during the annotation process. Work items where there was disagreement (no consensus) among annotators can be prioritized for review, because those work items are likely to have errors or illustrate ambiguity in the guidelines. In addition, annotators whose work items differ the most from the others' can be prioritized for review because they are likely to have more erroneous annotations and potentially require further training. One commonly used agreement metric is Krippendorff's alpha. A good resource for measures of inter-annotator agreement is *Handbook of Inter-rater Reliability*, by Kilen Li Gwet.

**Random QA** can be used to assess quality during the annotation process as well. This entails randomly sampling work items from each annotator to scrutinize for quality. The size of the sample for random QA will depend on the annotation project managers' availability to review each work item individually and determine whether the annotations therein are valid and reliable.

Findings from any of the above quality assurance methods can be used to evaluate and rank annotator performance, as well as to determine whether additional training or updates to the guidelines are needed. As part of the quality assurance process, the annotation tool should permit annotation project managers to remove or replace erroneous annotations so they are not included in the final data set submitted for downstream use. In addition, the tool should facilitate the communication of written feedback to the annotators with respect to erroneous work items, as well as real-time tracking of quality, timeliness and productivity statistics (see below).

At the outset of the annotation project, stakeholders should agree on a certain level or percentage of work items that should be audited for quality assurance. However, as the project progresses, the frequency with which annotations are assessed – whether automatically through "gold" tasks or manually sampled through random QA – may be influenced by the project's workforce and the quality of the annotations it has produced to date. For example, the Crowd may require the same high level of quality assessment throughout the entirety of the annotation process, regardless of annotation quality, due to high annotator turnover. However, vendors and full-time employees may not require the same level of quality assessment throughout the process given there is usually less annotator attrition; instead, the quality of the annotations themselves may indicate whether a higher or lower level of quality assessment is needed as the project progresses. For those workforces, if the error rate is higher than expected, it may make sense to review a larger sample of work items than originally anticipated so that more feedback can be provided to the annotators and more annotations can be remediated. Conversely, if the error rate is lower than expected, it may not be worth the time and resources to continue performing the same amount of review.

## Explicitly Address Annotator Collaboration

Depending on the workforce, project, and quality assurance method, collaboration among annotators may or may not be desirable. For example, annotators should not collaborate if annotation redundancy will be leveraged for quality assurance, as collaboration will taint the results. Similarly, annotators should not collaborate on projects that require subjective judgment calls because collaboration will introduce bias. In addition, it may be undesirable for annotators to collaborate when they are all at the same experience and skill level, as collaborating – instead of seeking clarification from annotation project managers – may perpetuate misunderstanding of the guidelines and lead to inconsistent or inaccurate annotations.

By contrast, annotator collaboration may be desirable for objective annotation projects where there are high-performing, experienced annotators who can help train newer annotators on how to apply the guidelines.

In all cases, it is important to include explicit instructions in the guidelines about whether collaboration among annotators is permitted. When collaboration is undesirable, the workforce's ability to collaborate can be mitigated by ensuring that the work items in each annotator's work queue are ordered differently so that the annotators are never working on the same work items at the same time.

## Provide the Workforce with Written Feedback for Erroneous Work Items

Providing the workforce with written feedback as part of the quality assurance process is critical to the success of any annotation project. Feedback should be aimed at increasing the annotators' understanding of the guidelines, so that more accurate and consistent annotations can be achieved going forward. To that end, feedback should consist of copies of erroneous work items, as well as clear, written explanations of the errors. This should include references to specific rules or examples from the guidelines, in addition to how they are applicable. Providing feedback in written form ensures that all annotators receive the same guidance simultaneously, which promotes annotation consistency and gives annotation project managers the ability to easily track when the guidance was provided.

Feedback should always be consistent with the guidelines. If a specific data scenario calls for feedback that is not covered by or departs from the guidelines, consideration should be given to whether an update or modification to the guidelines is warranted (see **Section VIII**).

To reduce the need for future remediation and optimize the project's failure time, feedback should be given as promptly as possible. Tracking when specific feedback was given can allow annotation project managers to determine whether feedback has been effective in eliciting improvements to annotation quality and consistency or whether changes to the guidelines, project, and/or workforce need to be considered.

## Promptly Respond to Questions from the Workforce

For objective annotation projects where annotation redundancy is not being employed, the workforce should be encouraged to escalate questions to annotation project managers when there is any confusion or uncertainty about how to apply the guidelines to specific work items. Annotation project managers should simplify the question escalation process by communicating directly with annotators, setting expectations as to how long it will take for annotators to receive responses to questions (e.g., within 24 hours), and ensuring that follow-up communication with annotators meets those expectations. When annotators are awaiting responses from annotation project managers, they should be instructed to "pend" (or "skip") the relevant work item so their workflow is not disrupted and they can continue to produce annotations. If the annotation tool does not allow annotators to "pend" work items, annotation project managers should endeavor to respond to questions in real-time so the workforce's productivity and timeliness are minimally impacted by any waiting time. Like feedback for erroneous work items, responses to questions should be provided in writing so that all annotators receive the same guidance simultaneously, which promotes annotation consistency and gives annotation project managers the ability to easily track when the guidance was provided.

Annotators should not be encouraged to raise questions for projects that require annotation redundancy or subjective judgment calls, because the additional guidance will skew the results. In addition, raising questions may not be desirable if the annotation tool does not allow the annotators to "pend" work items and annotation project managers are unable to respond to questions in real-time, because the workforce's productivity and timeliness may be significantly impacted.

## Track Quality, Timeliness, and Productivity Statistics

Annotation project managers should measure the workforce's quality, timeliness, and productivity. Timeliness (time per work item) and productivity (total number of work items completed) statistics provide insight into the rate of annotation collection, which informs the project's status and impacts projected timelines. Quality statistics provide insight into the overall accuracy and consistency of the annotations that are being produced by the workforce.

The annotation tool should automatically calculate and record these statistics in real-time on an individual annotator, as well as an aggregate basis. If that is not possible, timeliness, productivity, and quality data should be readily available for export from the tool so these statistics can be calculated and stored on another platform. For example, there should be data about work items that were marked incorrect by annotation project managers so that quality scores can be computed. There should also be data from time-stamped audit trails for each work item so that timeliness and productivity can be measured.

The annotation project manager should share these statistics with engineering and product stakeholders. It is at the discretion of the annotation project manager whether the statistics should also be shared with the annotators, given that full transparency may or may not positively affect the behavior of the workforce.

Real-time statistics can be leveraged to provide more focused quality assurance and feedback to the workforce. For example, individual annotator quality statistics can be used to determine whether the work items of specific annotators should be targeted for more quality assurance review going forward. In addition, individual annotator productivity/timeliness statistics can provide insight on possible root causes of poor quality annotations, such as inattentiveness due to speed.

Aggregate statistics can be used to monitor the project's status, as well as to confirm that the workforce is capable of following the guidelines and appropriate for the project. If the relevant quality, productivity, and timeliness targets have not been met, this data should be used to gain strategic insight on whether the guidelines, project, and/or workforce should be reconsidered and potentially changed.

When employing vendors as the workforce, aggregate statistics may also be required to determine the vendors' compliance with the quality, productivity, and timeliness requirements set forth in Service Level Agreements.

With the Crowd and some vendors, timeliness, productivity, and quality statistics are typically used to set a payment rate, award bonuses, or impose monetary penalties. If annotations are proceeding slower or quicker than initially estimated, financial benchmarks may need to be adjusted. It is important to ensure that annotators are not incentivized or pressured to "rush" annotations in order to achieve a higher payment rate or secure a living wage. Quality must always be the primary metric for any workforce.

## Ensure Technical Support for the Annotation Tool

Prior to engaging the workforce, the stakeholders should determine who will be responsible for troubleshooting any technical issues that may arise with the annotation tool during the project. It may be helpful to outline different scenarios ahead of time (e.g., server or tool unexpectedly going offline) to decide how those issues should be escalated and addressed, especially if the workforce will be using the tool outside of the annotation project managers' normal business hours. Technical support should be ensured so issues can be addressed quickly to minimize the disruption of annotators' workflows and prevent project delays.

# XI.  After Annotation is Complete

## Summary

- It is important **to assess overall annotation quality after the project has been completed**, so that the validity and reliability of the annotations can be confirmed before they are submitted for downstream use.

- If additional annotations are required to achieve the project's end goal after the initial project run has been completed, it is important to **consider strategic adjustments** to the project's definition, training process, and workforce, so the next round of annotation collection can be more efficient.

- It is important to implement processes to **detect data drift and anomalies** that may require additional annotations.

## Assess Overall Annotation Quality After the Project Has Been Completed

Although it is important to consistently monitor the quality of the annotations produced by the workforce throughout the project so that feedback can be provided, the overall quality of the annotations should also be assessed after the project is complete.

When possible, consensus annotations should be provided as part of the final project output. There are many methods for determining consensus annotations, including majority voting and variations on trust/reliability weighted voting (see Dirk Hovy, Taylor Berg-Kirkpatrick, Ashish Vaswani, and Eduard Hovy. 2013. "Learning whom to trust with MACE." In Proceedings of the 2013 Conference of the Northn American Chapter of the Association for Computational Linguistics: Human Language Technologies. Atlanta, Georgia, USA, NAACL-HLT-13, pages 1120-1130). For example, stakeholders may consider:

- giving greater weight to annotations produced by annotators identified as high-performing;

- eliminating annotations produced by annotators identified as low-performing; and/or

- funneling low-consensus annotations into an escalation queue for remediation by annotation project managers.

It should be emphasized that these methods are designed to identify potentially erroneous annotations for removal or remediation, and should not be used to exclude data points that were difficult for the workforce to annotate, as that would skew the data and have a negative impact on the use of the data (e.g., for machine learning).

In addition, annotation project managers may consider performing a manual review of a percentage of the final annotations to confirm their validity and reliability, as well as searching for patterns in the data set and/or quality metrics to identify potential bias.

## Consider Strategic Adjustments if Additional Annotations Are Needed

After the initial project run has been completed, stakeholders should determine whether more annotations than originally anticipated are needed for downstream use to achieve the project's end goal. Given that additional annotations require more resources in terms of time, budget, and effort, further annotation collection should be carefully considered and approached strategically.

For example, stakeholders should leverage insights gained from the quality assurance process, as well as accuracy, timeliness, and productivity data to determine whether efficiencies can be gained for the next round of annotation collection. Considerations should include whether the tool, guidelines, and training process could be improved; whether a change in workforce could be more cost-effective; and whether further microtasking could lead to potentially higher quality and reusable data.

If the annotations are being used for a machine learning model, the model's performance can be used to estimate how many additional annotations may be needed and whether certain types of data/examples should be targeted for additional annotation. For example, it may be beneficial to target:

- examples where the model was confused;

- examples where the model differed from the evaluation data; and/or

- examples that meet different probability thresholds.

In all cases, stakeholders should calculate or derive a return on investment that details how many additional annotations (and associated annotator-hours) will lead to measurable improvements in performance metrics, so that a specific budget and resource allocation can be justified.

In addition, stakeholders should consider whether or not the current set of annotations is adequate to produce a minimum viable product, with the expectation that additional annotations may be collected at a later date or, if a machine learning model is being leveraged, acquired through active learning.

## Detect Data Drift and Anomalies

All annotation projects, regardless of their downstream use, will require some budget and resource allocation in perpetuity for monitoring and quality assurance. In particular, it is important to implement processes to detect the following two phenomena through ongoing annotation:

- *Data drift* – Data drift occurs when the distribution of annotation labels, or features of the data, change slowly over time. Data drift can lead to increasing error rates for machine learning models or rule-based systems. There is no static data: ongoing annotation is necessary to adapt downstream models/solutions as data drift occurs.

- *Anomalies* – While data drift refers to slow changes in data, anomalies are step functions – sudden (and typically temporary) changes in data due to exogenous events. For example, in 2019-20, the COVID-19 pandemic led to anomalies in many naturally occurring data sets. It is important to have procedures in place to detect anomalies. When anomalies occur, it may be necessary to shift from automated solutions to human-based workflows, or to temporarily ramp up annotations so downstream models/solutions can be quickly adapted.

# XII. Conclusions

Data annotation is the invisible workhorse of machine learning. Without annotated data, none of the machine learning solutions we rely on today would be available. Importantly, the quality, coverage, and accuracy of annotated data sets determine the quality, coverage, and accuracy of the machine learning solution; just as with traditional software engineering, *garbage in, garbage out.* Yet, machine learning engineers and data scientists typically are taught about every aspect of the machine learning workflow *except* how to design and manage annotation projects.

This practical guide includes a lot of advice specific to annotation projects, as well as key elements that ensure the success of any project:

- Engage all stakeholders

- Define your goal clearly

- Plan your roadmap

- *Communicate* early and often

There is a rapidly growing research discipline, *human computation,* that is focused on how to optimize the interaction of humans and computers in order to obtain accurate and timely annotated data. For readers who want to learn more about the science of data annotation and advances in annotation for machine learning, we advise checking out the [Association for the Advancement of Artificial Intelligence's (AAAI) Conference on Human Computation and Crowdsourcing (HCOMP)](#), as well as papers on data annotation, crowdsourcing, and human computation published at major machine learning and NLP conferences, such as AAAI, KDD, and NeurIPS, in addition to those held annually by the Association for Computational Linguistics (ACL).

# About the Authors

**Tina Tseng** has been a legal analyst at Bloomberg since 2008. For the last 10 years, she has worked on developing legal research tools that leverage natural language processing and machine learning, including Bloomberg Law's Points of Law product, which received the 2018 New Product Award from the American Association of Law Libraries. Tina earned her Bachelor of Arts degree from Cornell University and her Juris Doctor from Georgetown University Law Center. Tina can be contacted at ttseng8@bloomberg.net.

**Amanda Stent** is a NLP architect in the Data Science Group in the Office of the CTO at Bloomberg. Previously, she was a director of research and principal research scientist at Yahoo Labs, a principal member of technical staff at AT&T Labs – Research, and an associate professor in the Computer Science Department at Stony Brook University. Her research interests center on natural language processing and its applications, particularly topics related to text analytics, discourse, dialog, and natural language generation. She holds a Ph.D. in computer science from the University of Rochester. She is co-editor of the book "Natural Language Generation in Interactive Systems" (Cambridge University Press), has authored over 100 papers on natural language processing, and is co-inventor on over 30 patents and patent applications. She is one of the rotating editors of the journal *Dialogue & Discourse*. Amanda can be contacted at astent@bloomberg.net.

**Domenic Maida** is the head of Bloomberg's Global Data group, which gathers, analyzes, and curates timely, comprehensive, and high-quality datasets that are foundational to the analytics and tools Bloomberg provides through its products, including the Bloomberg Terminal. In this role, he oversees the teams responsible for acquiring data and managing relationships with more than 350 exchanges, regulators, and trading venues. He started his career at Bloomberg as a software engineer in 1995, later led the software infrastructure and applications teams, and then became the global manager of Research and Development. From 2008 to 2012, he held the position of global head of Bloomberg's Terminal business. He graduated from Johns Hopkins University, where he earned a Bachelor of Science in mechanical and electrical engineering.